\documentclass[final,3p,times]{elsarticle}  

\usepackage{graphics}
\usepackage{graphicx}

\usepackage{amssymb}
\usepackage{physics}
\usepackage{slashed}
\usepackage{bm}
\usepackage{url}
\usepackage{booktabs}
\usepackage{color}
\usepackage{mmacells}
\usepackage{subcaption}
\usepackage{extarrows}
\usepackage{fancyvrb}
\usepackage{xcolor}
\usepackage{multirow}

\allowdisplaybreaks[1]

\usepackage{amsthm}

\usepackage{lineno}

\newcounter{bla}

\newcommand{\blue}[1]{{\color{blue}{#1}}} 

\newcommand{\Mathematica}{\texttt{Mathematica}}
\newcommand{\FeynRules}{\texttt{FeynRules}}
\newcommand{\FeynArts}{\texttt{FeynArts}}
\newcommand{\FeynCalc}{\texttt{FeynCalc}}
\newcommand{\MadGraph}{\texttt{MadGraph}}
\newcommand{\WHIZARD}{\texttt{WHIZARD}}

\journal{Computer Physics Communications}

\begin{document}

\begin{frontmatter}



\title{Toponium: Implementation of a toponium model in FeynRules}


\author[a]{Jing-Hang~Fu}
\author[a]{Yu-Jie~Zhang}
\author[b]{Guang-Zhi Xu}
\author[c]{Kui-Yong Liu\corref{author}}

\cortext[author] {Corresponding author.\\\textit{E-mail address:} liukuiyong@lnu.edu.cn}
\address[a]{School of Physics, Beihang University, Beijing, China.}
\address[b]{School of Physics, Liaoning University, Shenyang, China.}
\address[c]{School of Physics, Liaoning Normal University, Dalian, China.}

\begin{abstract}





Toponium---a bound state of the top-antitop pair ($t\bar{t}$)---emerges as the  smallest and simplest hadronic system in QCD, with an ultrashort lifetime ($\tau_t \sim 2.5\times 10^{-25}$~s) and a femtometer-scale Bohr radius ($r_{\text{Bohr}} \sim 7\times 10^{-18}$~m). We present a computational framework extending the Standard Model (SM) with two S-wave toponium states: a spin-singlet $\eta_t$ ($J^{PC}=0^{-+}$) and a spin-triplet $J_t$ ($J^{PC}=1^{--}$). Using nonrelativistic QCD (NRQCD) and a Coulomb potential, we derived couplings to SM particles (gluons, electroweak bosons, Higgs boson, and fermion pairs) and implemented the Lagrangian in FeynRules, generating FeynArts, MadGraph, and WHIZARD models for collider simulations.  Key results include dominant decay channels ($\eta_t \to gg/ZH$, $J_t \to W^+W^-/b\bar{b}$) and leading order (LO) cross sections for  $pp \to \eta_t(nS) \to {\rm non-}t\bar{t}$ ({66 fb} at 13 TeV). The model avoids double-counting artifacts by excluding direct $t\bar{t}$ couplings,  thereby ensuring consistency with perturbative QCD. This work establishes a complete pipeline for precision toponium studies, bridging NRQCD, collider phenomenology, and tests of SM validity at future lepton colliders  (e.g., CEPC, FCC-ee, muon colliders) and the LHC.  This provides the first publicly available UFO model for toponium, enabling direct integration with MadGraph and WHIZARD for simulations.


\noindent \textbf{PROGRAM SUMMARY}

\begin{small}
\noindent
{\em Program Title: Toponium}                                          \\
{\em CPC Library link to program files:} \blue{(to be added by Technical Editor)} \\
{\em Developer's repository link:} \url{https://github.com/fujinghang/Toponium} \\
{\em Licensing provisions(please choose one):} BSD 3-clause  \\
{\em Programming language: \tt{Wolfram Mathematica}}                                   \\
{\em Nature of problem:} Toponium---the smallest bound state and simplest hadron composed of a top-antitop pair ($t\bar{t}$)---plays a pivotal role in testing quantum mechanics and probing SM. The framework is critical for disentangling toponium signals from $t\bar{t}$ continuum backgrounds at the LHC and future lepton colliders (CEPC, FCC-ee, and muon colliders), where precise studies of threshold dynamics and quantum entanglement require robust theoretical tools.   
\\
{\em Solution method:} We extend the SM with two S-wave toponium states: a spin-singlet $\eta_t$ ($J^{PC}=0^{-+}$) and a spin-triplet $J_t$ ($J^{PC}=1^{--}$). The framework integrates nonrelativistic QCD (NRQCD) and Coulomb potential dynamics to derive the couplings between toponium and SM particles, including gluons, electroweak bosons ($W^\pm$, $Z$, $\gamma$), Higgs fields, and fermion pairs.  The theoretical framework is implemented computationally by the  \Mathematica~[1] package \FeynRules~[2] and \FeynCalc~[3], which produce  both \FeynArts~[4] model files for diagrammatic calculations and UFO formats for collider simulations in \MadGraph~[5] and \WHIZARD~[6]. This creates the first complete computational pipeline for precision studies of toponium production, decay spectra, and threshold dynamics at high-energy colliders. 
\\
{\em Additional comments including restrictions and unusual features:} The model intentionally excludes direct couplings between the toponium states ($\eta_t$, $J_t$) and the top quark pair $t\bar{t}$, since these would lead to fundamental inconsistencies when incorporated into higher-order QCD calculations, in particular through problematic double-counting of the top quark's weak decay channels.
   \\
   

\end{small}
   \end{abstract}
\end{frontmatter}


\section{Introduction}

The bound state of toponium ($t\bar{t}$), formed via strong interactions, is the smallest known quantum mechanical bound state~\cite{Fu:2024bki}, with a Bohr radius of $7.4 \times 10^{-18}$~m and an extremely short lifetime of $2.5 \times 10^{-25}$~s. These characteristics make toponium a distinctive platform for investigating fundamental quantum mechanical phenomena, such as entanglement~\cite{Maltoni:2024csn,ATLAS:2023fsd,Aguilar-Saavedra:2024fig,Han:2024ugl}, and perturbative confinement mechanisms, free from the complications of hadronization that obscure lighter quarkonium.

Toponium dynamics also provides a sensitive probe of the SM, particularly through its dependence on the Higgs-top Yukawa coupling~\cite{Feigenbaum:1990sr}, electroweak symmetry~\cite{Sehgal:1980ja}, and vacuum~\cite{Voloshin:1978hc}. Recently, the CMS collaboration observed a pseudoscalar excess at the top quark pair production threshold~\cite{CMS:2025kzt}. Precision measurements of its production and decay at future colliders, such as the Circular Electron Positron Collider (CEPC)~\cite{CEPCStudyGroup:2018ghi,CEPCStudyGroup:2023quu}, the Future Circular Lepton Collider (FCC-ee)~\cite{FCC:2018evy}, and the muon colliders~\cite{Accettura:2023ked,Han:2024gan,InternationalMuonCollider:2024jyv,Han:2025wdy}, could unveil deviations in \( t\bar{t} \) interactions or Higgs-mediated processes, offering indirect hints of physics beyond the SM.

From an experimental perspective, the development of FeynRules/UFO models for toponium allows systematic simulations of its collider signatures, such as \( J_t \to W^+W^-/b\bar{b} \) and \( \eta_t \to ZH/W^+W^- \). These predictions are critical for distinguishing toponium signals from \( t\bar{t} \) continuum backgrounds and exotic resonances at the Large Hadron Collider (LHC) and next-generation colliders. Together, these efforts provide a fundamental pipeline for advancing precision top quark physics, refining new physics searches, and exploring quantum chromodynamics in relativistic extremes previously inaccessible to theoretical and experimental scrutiny. 
While previous  studies~\cite{Fu:2024bki,Fuks:2021xje,Aguilar-Saavedra:2024mnm,Fuks:2024yjj,jiang2024studytoponiumspectrumassociated,Francener:2025tor} have explored toponium signatures, this work provides the first publicly available UFO model for toponium to our knowledge, enabling direct integration with MadGraph and WHIZARD for simulations. This advance addresses a critical gap in collider-based toponium studies, as earlier approaches have relied on ad hoc implementations that are incompatible with modern event generators.

This study focuses on developing a computational framework to model the production and decay processes of toponium. We utilize Mathematica-based packages, including \FeynArts~\cite{Hahn:2000kx}, \FeynCalc~\cite{Mertig:1990an,Shtabovenko:2020gxv}, and \FeynRules~\cite{Christensen:2008py,Alloul:2013bka}. \FeynArts~is employed to generate Feynman diagrams, while \FeynCalc~is used to calculate the corresponding Feynman amplitudes. To incorporate the effects of toponium bound states, we combine these results with nonrelativistic quantum chromodynamics (NRQCD) \cite{Bodwin:1994jh,Brambilla:2004jw}, which provides a systematic framework for  describing the production and decay of heavy quarkonium. Within the NRQCD approach, we calculate the coupling vertices involving toponium, including both the S wave spin singlet ($\eta_t$) and spin triplet ($J_t$) ground states, and derive the relevant Lagrangian for toponium interactions.

We implement the derived Lagrangian in \FeynRules, by extending the SM to include toponium states. This allows us to generate the \FeynRules~model file, which serves as the foundation for further phenomenological studies. From this model file, we produce both \FeynArts~and UFO model files. The UFO model, in particular, is designed for compatibility with modern event generators such as \MadGraph~\cite{Alwall:2014hca} and \WHIZARD~\cite{Kilian:2007gr}, enabling efficient simulation of toponium production and decay processes in high-energy experiments. This comprehensive framework not only facilitates the study of toponium properties but also provides a robust tool for exploring new physics scenarios beyond the SM.

\section{\texorpdfstring{Lagrangian of $\eta_t$ and $J_t$}{Lagrangian of Et and Jt}}

\subsection{\texorpdfstring{The bound states of $\eta_t$ and $J_t$ }{The bound states of Et and Jt}}

In the context of nonrelativistic quantum mechanics, the state vector of the bound state \( B \) composed of \( t\bar{t} \) can be described by the \( t\bar{t} \) state vector and the zero-point wave function~\cite{Bodwin:1994jh,Peskin:1995ev}:
\begin{equation}
\begin{aligned}
\ket{B} = \sum_{s_{1z}, s_{2z}, S_{z}, L_{z}, i, j} \braket{s_{1}, s_{1z}; s_{2}, s_{2z}}{S, S_{z}} \braket{S, S_{z}; L, L_{z}}{J, J_{z}} \braket{3i; 3j}{1}  \times \int \frac{\dd^3{k}}{(2\pi)^{3/2}} \Tilde{\psi}(\bm{k}) \ket{t_{1i} \bar{t}_{2j}},
\end{aligned}
\label{eq:state_vector}
\end{equation}
where $\ket{t_{1i} \bar{t}_{2j}}$ is the $t\bar{t}$ state vector, $\Tilde{\psi}(\bm{k})$ is the momentum-space wave function of $B$, $\bm{k}$ is the relative momentum of the $t\bar{t}$ pair, and the coefficients $\braket{s_{1}, s_{1z}; s_{2}, s_{2z}}{S, S_{z}}$, $\langle S, S_{z}; L, L_{z} | J, J_{z} \rangle$, and $\langle 3i; 3j | 1 \rangle$ represent the Clebsch-Gordan (CG) coefficients for spin-spin coupling, spin-orbit coupling, and color projection of $t \bar{t}$ to the bound state $B$, respectively.

For the decay process $B \to p_3 p_4$, by performing a Taylor expansion in $k$ and transforming the wave function to coordinate space, we obtain:
\begin{eqnarray}
{\cal M}(B\to p_3 p_4)
&=&\int \frac{\dd^3{k}}{(2\pi)^{3/2}}\Tilde{\psi}(\bm{k}){\cal M}(t_{1}\bar{t}_{2}\to p_3 p_4;\bm{k})\nonumber\\
&=&\int \frac{\dd^3{k}}{(2\pi)^{3/2}}\Tilde{\psi}(\bm{k})
\bigg(
{\cal M}(t_{1}\bar{t}_{2}\to p_3 p_4;\bm{0})+\nabla_{\bm{k}}{\cal M}(t_{1}\bar{t}_{2}\to p_3 p_4;\bm{k})\Big|_{\bm{k}=\bm{0}}\cdot\bm{k}
+\ldots
\bigg)\nonumber\\
&=&{\cal M}(t_{1}\bar{t}_{2}\to p_3 p_4;\bm{0})\psi(\bm{0})+\nabla_{\bm{k}}{\cal M}(t_{1}\bar{t}_{2}\to p_3 p_4;\bm{k})\Big|_{\bm{k}=\bm{0}}\cdot-i\nabla_{\bm{r}}\psi(\bm{r})\Big|_{\bm{r}=\bm{0}}
+\ldots
\end{eqnarray}
where $\psi(\bm{r})$ is the coordinate-space wave function of $B$. The S-wave corresponds to the first term, and the P-wave corresponds to the second term. The CG coefficients are included in the amplitude ${\cal M}(t_{1}\bar{t}_{2}\to p_3 p_4;\bm{k})$. Here, $\psi(\bm{0})$ is the value of the wave function at the origin, which can be obtained by solving the Schr\"{o}dinger equation. To project from $t\bar{t}$ to the bound state $B$, a projection operator $P$ is introduced, where $K$ is the four-momentum of $B$:
\begin{eqnarray}
P_{SS_z}(K,k)=\sum_{s_{1z},s_{2z}}\braket{s_{1},s_{1z};s_{2},s_{2z}}{S,S_{z}}u\left(\frac{K}{2}+k;s_1\right)\bar{v}\left(\frac{K}{2}-k;s_2\right).
\end{eqnarray}
For the relative momentum $k \ll m_t$, the S-wave projection operator only needs to be retained up to the LO in $k$.
\begin{eqnarray}
    P_{00}(K,0)&=&-\frac{1}{2\sqrt{2m_t}}(\slashed{K}+2m_t)\gamma_5,\nonumber\\
    P_{1S_z}(K,0)&=&\frac{1}{2\sqrt{2m_t}}(\slashed{K}+2m_t)\slashed{\epsilon}(S_z).
\end{eqnarray}
The first equation corresponds to the S-wave spin singlet ($\eta_t$, $J^{PC} = 0^{-+}$) and the spin triplet ($J_t$, $J^{PC} = 1^{--}$). Here, $\epsilon$ is the polarization vector of $J_t$, and its polarization sum relation is given by:
\begin{eqnarray}
\sum_{S_z}\epsilon_\mu^\ast(S_z)\epsilon_\nu(S_z)=-g_{\mu\nu}+\frac{K_\mu K_\nu}{(2m_t)^2}.
\end{eqnarray}

We adopt the Coulomb potential obtained from lattice QCD calculations~\cite{Kawanai:2013aca,Koma:2006fw} as the interaction potential for $t\bar{t}$:
\begin{eqnarray}
    V(r)=-\frac{\lambda}{r},
\end{eqnarray}
where  $r$ is the distance between $t$ and $\bar{t}$, and the coefficient {$\lambda=0.309$} is fitted from the $\eta_t$ cross section measured by CMS~\cite{CMS:2025kzt}.
The Cornell potential \(V(r)=-{\lambda}/{r}+\sigma r\) is reproduced in the lattice QCD calculations with $\sigma=0.206~{\rm GeV}^2$~\cite{Kawanai:2013aca,Koma:2006fw}.
The linear term $\sigma r$ in the potential can be neglected, as its contribution to the mass of the ground toponium states is 0.012 GeV. So the Coulomb potential is chosen for its dominance in the toponium system.

By solving the Schr\"{o}dinger equation, one can obtain the wave function at the origin,
\begin{eqnarray}
    \abs{\psi_{nS}(0)}^2&=&\frac{(\lambda m_t)^3}{8\pi n^3},
\end{eqnarray}
where $n=1,2,3,\dots$ is the principal quantum number of $J_t(nS)$ and $\eta_t(nS)$. And  $\eta_t$($J_t$) refers to  $\eta_t(1S)$($J_t(1S)$).

Then the decay amplitude of $B$ is given by:
\begin{eqnarray}
    {\cal M}(B\to p_3 p_4)
    ={\cal M}(t_1\bar{t}_2\to p_3 p_4;\bm{0})\sqrt{\frac{\lambda^3 m_t^3}{8\pi}}.
\end{eqnarray}
Here, the momentum of the top quark and the anti-top quark are equal and on-shell.
We can then use this amplitude to remove the outer legs and obtain the coupling vertices for $\eta_t$ and $J_t$, as well as the corresponding Lagrangian.

The dominant decay channel for $\eta_t$ and $J_t$, in terms of decay width, is the weak decay process $\eta_t/J_t \to t\bar{t} \to W^+bW^-\bar{b}$. However, if the weak decay channel of $t\bar{t}$ is included directly in the calculation, higher-order computations may lead to incorrect results due to improper treatment of intermediate states and their associated width effects. Therefore, we restrict our calculation to the annihilation decay channels of $\eta_t$ and $J_t$, excluding weak decay processes of $t\bar{t}$ such as $\eta_t/J_t \to W^+bW^-\bar{b}$~\cite{BuarqueFranzosi:2017jrj}. 

\subsection{\texorpdfstring{Lagrangian of $\eta_t$}{Lagrangian of Et}}

In addition to the weak decay channel $\eta_t \to t\bar{t}$, the annihilation decay channels of $\eta_t$ include processes such as $gg$, $HZ$, $W^+W^-$, $ZZ$, $ZA$, and $AA$. Representative Feynman diagrams for these channels are shown in Fig.~\ref{fig:EtDecayFeynmanDiagram}. The decay amplitudes for these processes are calculated within the framework of NRQCD:
\begin{figure}[htbp]
    \centering
    \begin{subfigure}[b]{0.3\textwidth}
        \includegraphics[width=\textwidth]{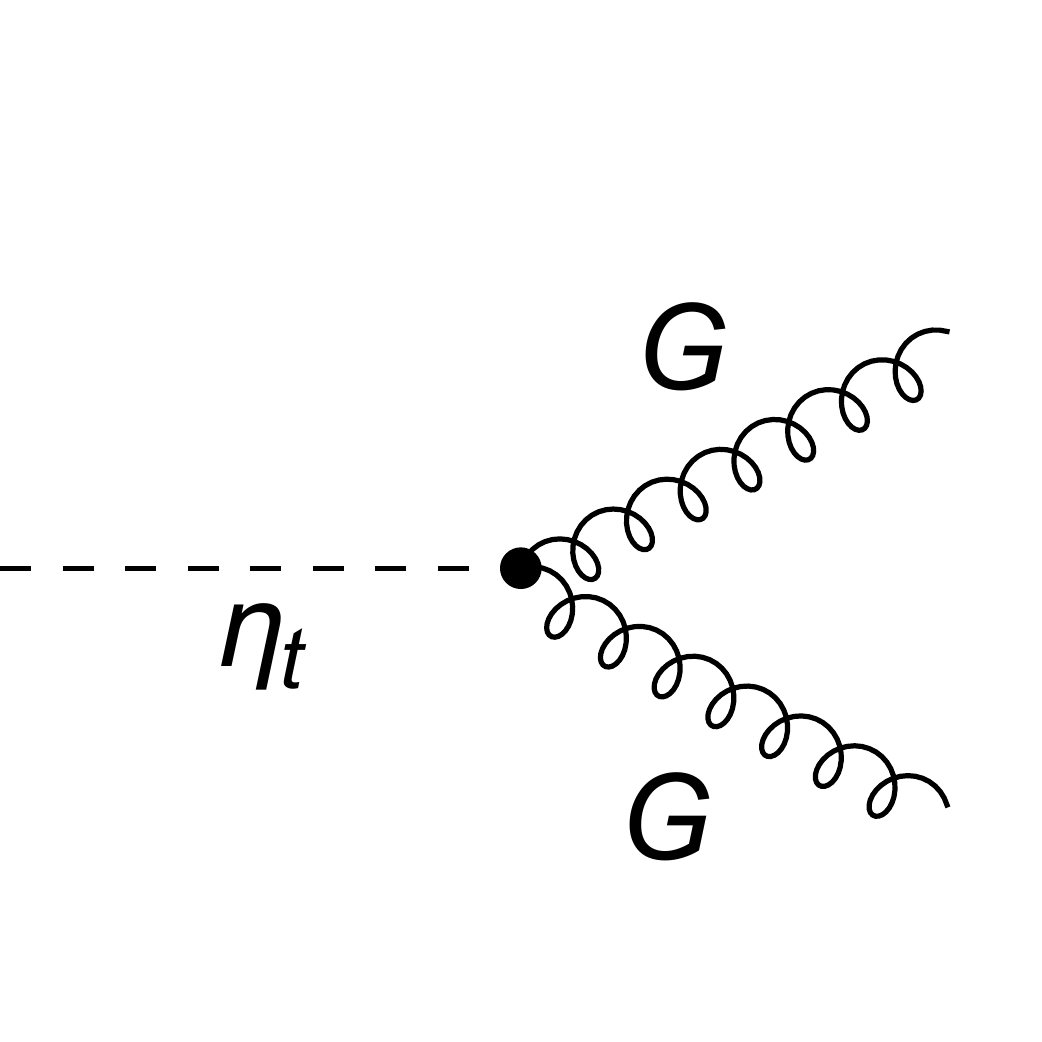}
    \end{subfigure}
    \begin{subfigure}[b]{0.3\textwidth}
        \includegraphics[width=\textwidth]{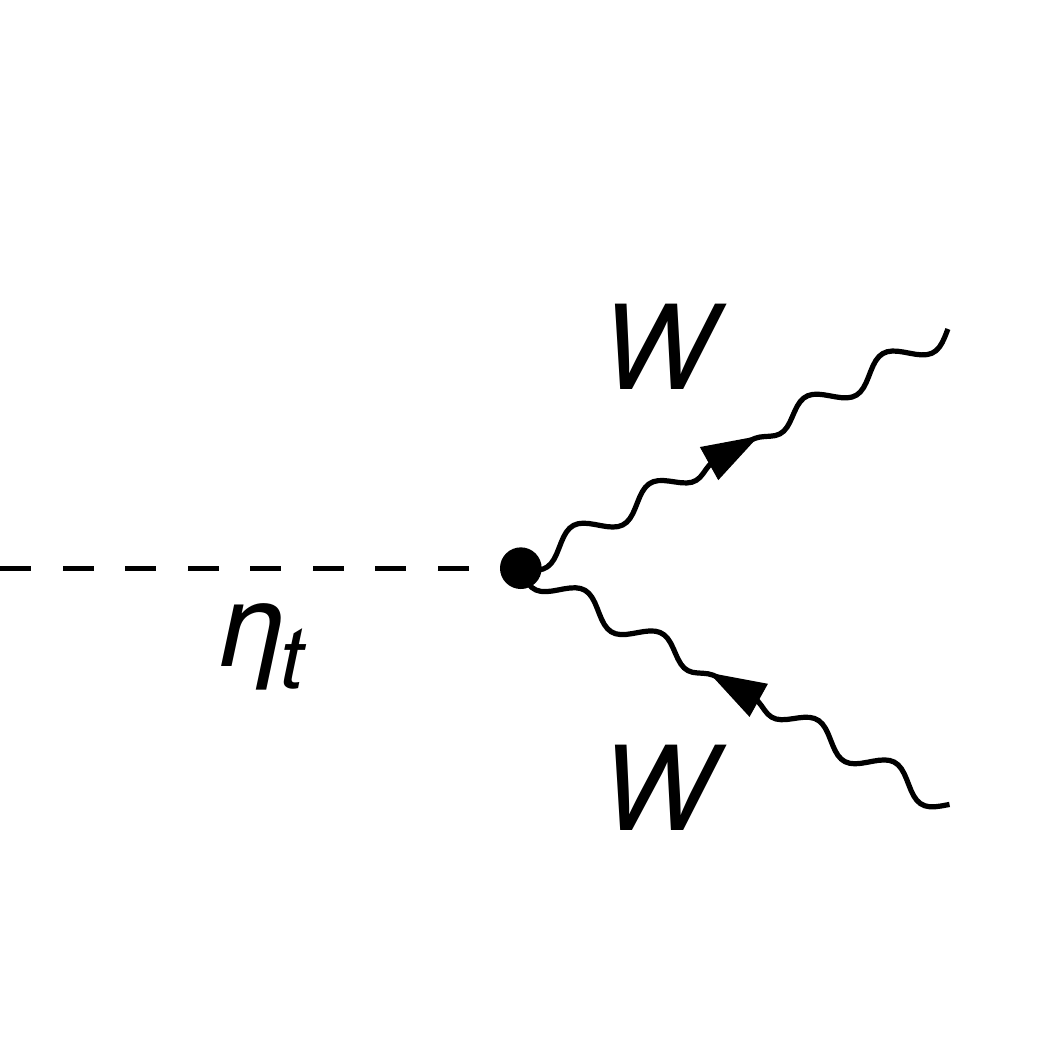}
    \end{subfigure}
    \begin{subfigure}[b]{0.3\textwidth}
        \includegraphics[width=\textwidth]{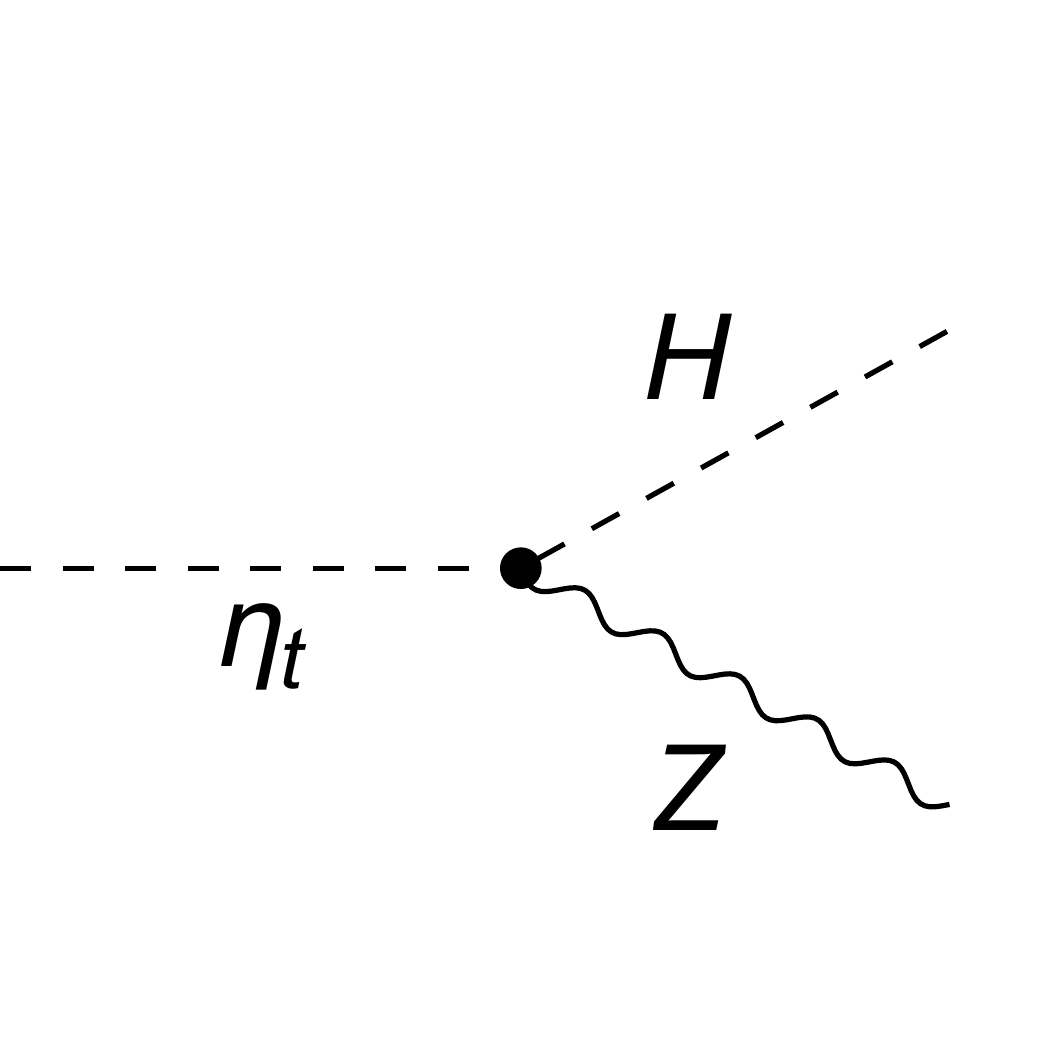}
    \end{subfigure}
    \caption{The three dominant Feynman diagrams of the decay of $\eta_t$.}
    \label{fig:EtDecayFeynmanDiagram}
\end{figure}
\begin{eqnarray}
    {\cal M}(\eta_t \to g_{1}g_{2})&=&4i g_{\eta_t gg}~\delta_{c_1,c_2}
    \epsilon_{\mu \nu \rho \sigma}k_{g_1}^\mu k_{g_2}^\nu \varepsilon^{*\rho}(k_{g_1}) \varepsilon^{*\sigma}(k_{g_2}),\nonumber\\
    {\cal M}(\eta_t \to H Z)&=& g_{\eta_t HZ}~m_Z^2~k_H\cdot \varepsilon^*(k_Z), \nonumber\\
    {\cal M}(\eta_t \to W^+ W^-)&=& 2i g_{\eta_t WW}~\epsilon_{\mu \nu \rho \sigma}k_{W^+}^\mu k_{W^-}^\nu \varepsilon^{*\rho}(k_{W^+}) \varepsilon^{*\sigma}(k_{W^-}), \nonumber\\
    {\cal M}(\eta_t \to Z_1 Z_2)&=& 4i g_{\eta_t ZZ}~\epsilon_{\mu \nu \rho \sigma}k_{Z_1}^\mu k_{Z_2}^\nu \varepsilon^{*\rho}(k_{Z_1}) \varepsilon^{*\sigma}(k_{Z_2}), \nonumber\\
    {\cal M}(\eta_t \to ZA)&=& 2i g_{\eta_t ZA}~\epsilon_{\mu \nu \rho \sigma}k_{Z}^\mu k_{A}^\nu \varepsilon^{*\rho}(k_Z) \varepsilon^{*\sigma}(k_A), \nonumber\\
    {\cal M}(\eta_t \to A_1 A_2)&=&4i g_{\eta_t AA}~\epsilon_{\mu \nu \rho \sigma}k_{A_1}^\mu k_{A_2}^\nu \varepsilon^{*\rho}(k_{A_1}) \varepsilon^{*\sigma}(k_{A_2}).
\end{eqnarray}
Here, $c_1$ and $c_2$ denote the color indices for the two gluons, respectively.  The corresponding coupling constants are given by:
\begin{eqnarray}
    g_{\eta_t gg}&=& \sqrt{\frac{\pi \lambda ^3}{3}}\frac{\alpha_{s}}{4m_t}, \nonumber\\
    g_{\eta_t HZ}&=& -\sqrt{\frac{\pi \lambda ^3}{3}}\frac{3\alpha m_t c_{\rm W}}{2 s_{\rm W}^2 m_W^3}, \nonumber\\
    g_{\eta_t WW}&=& \sqrt{\frac{\pi \lambda ^3}{3}}\frac{3\alpha m_t}{4s_{\rm W}^2 (m_t^2-m_W^2)}, \nonumber\\
    g_{\eta_t ZZ}&=& \sqrt{\frac{\pi \lambda ^3}{3}} \frac{\alpha  m_t \left(8 s_{\rm W}^2(4 s_{\rm W}^2-3)+9\right)}{24 c_{\rm W}^2 s_{\rm W}^2 \left(2 m_t^2-m_Z^2\right)}, \nonumber\\
    g_{\eta_t ZA}&=& \sqrt{\frac{\pi \lambda ^3}{3}} \frac{2 \alpha  m_t \left(8 s_{\rm W}^2-3\right)}{3 c_{\rm W} s_{\rm W} \left(4 m_t^2-m_Z^2\right)}, \nonumber\\
    g_{\eta_t AA}&=& \sqrt{\frac{\pi \lambda ^3}{3}} \frac{2 \alpha }{3 m_t}.
\end{eqnarray}

The Lagrangian can be derived as follows:
\begin{eqnarray}
{\cal L}_{\eta_t}&=&\frac{1}{2}(\partial \eta_t)^2-\frac{1}{2}m_{\eta_t}\eta_t^2, \nonumber\\
{\cal L}_{\eta_t gg}&=& g_{\eta_t gg}G_{\mu \nu}^a \tilde{G}^{a \mu \nu}\eta_t, \nonumber\\
{\cal L}_{\eta_t HZ}&=& g_{\eta_t HZ}Z^{\mu\nu}(\partial_\mu H)(\partial_\nu \eta_t), \nonumber\\
{\cal L}_{\eta_t WW}&=& g_{\eta_t WW}W_{\mu \nu} \tilde{W}^{\dagger \mu \nu}\eta_t,  \nonumber\\
{\cal L}_{\eta_t ZZ}&=& g_{\eta_t ZZ}Z_{\mu \nu} \tilde{Z}^{\dagger \mu \nu}\eta_t, \nonumber\\
{\cal L}_{\eta_t ZA}&=& g_{\eta_t ZA}Z_{\mu \nu} \tilde{A}^{\dagger \mu \nu}\eta_t, \nonumber\\
{\cal L}_{\eta_t AA}&=& g_{\eta_t AA}A_{\mu \nu} \tilde{A}^{\dagger \mu \nu}\eta_t.
\end{eqnarray}
Here, the first term is the kinetic term for $\eta_t$ and the mass term, and the remaining terms describe the interactions. $G^{a}_{\mu \nu}$ and $W_{\mu \nu}$ are the field strength tensors for the gluon and $W$ boson, respectively ($G^{a}_{\mu \nu} = \partial_\mu G^{a}_\nu-\partial_\nu G^{a}_\mu + g_s f^a_{bc} G^{b}_\mu G^{c}_\nu,W_{\mu \nu} = \partial_\mu W_\nu-\partial_\nu W_\mu,Z_{\mu \nu} = \partial_\mu Z_\nu-\partial_\nu Z_\mu,A_{\mu \nu} = \partial_\mu A_\nu-\partial_\nu A_\mu$), while $\tilde{X}_{\mu \nu}~(X=G,W,Z,A)$ is dual field strength tensors ($\tilde{X}_{\mu \nu} = \frac{1}{2}\varepsilon_{\mu \nu \rho \sigma} X^{\rho \sigma}$).

\subsection{\texorpdfstring{Lagrangian of $J_t$}{Lagrangian of Jt}}

The annihilation decay channels of $J_t$ comprise processes such as $\nu\bar{\nu}$, $l^+l^-$, $u\bar{u}$, $d\bar{d}$, $s\bar{s}$, $c\bar{c}$, $b\bar{b}$, $HZ$, $HA$, $ZZ$, $ZA$, and $W^+W^-$. Representative Feynman diagrams for these channels are displayed in Fig.~\ref{fig:JtDecayFeynmanDiagram}. The decay amplitudes for these processes are calculated as follows:
\begin{figure}[htbp]
    \centering
    \begin{subfigure}[b]{0.3\textwidth}
        \includegraphics[width=\textwidth]{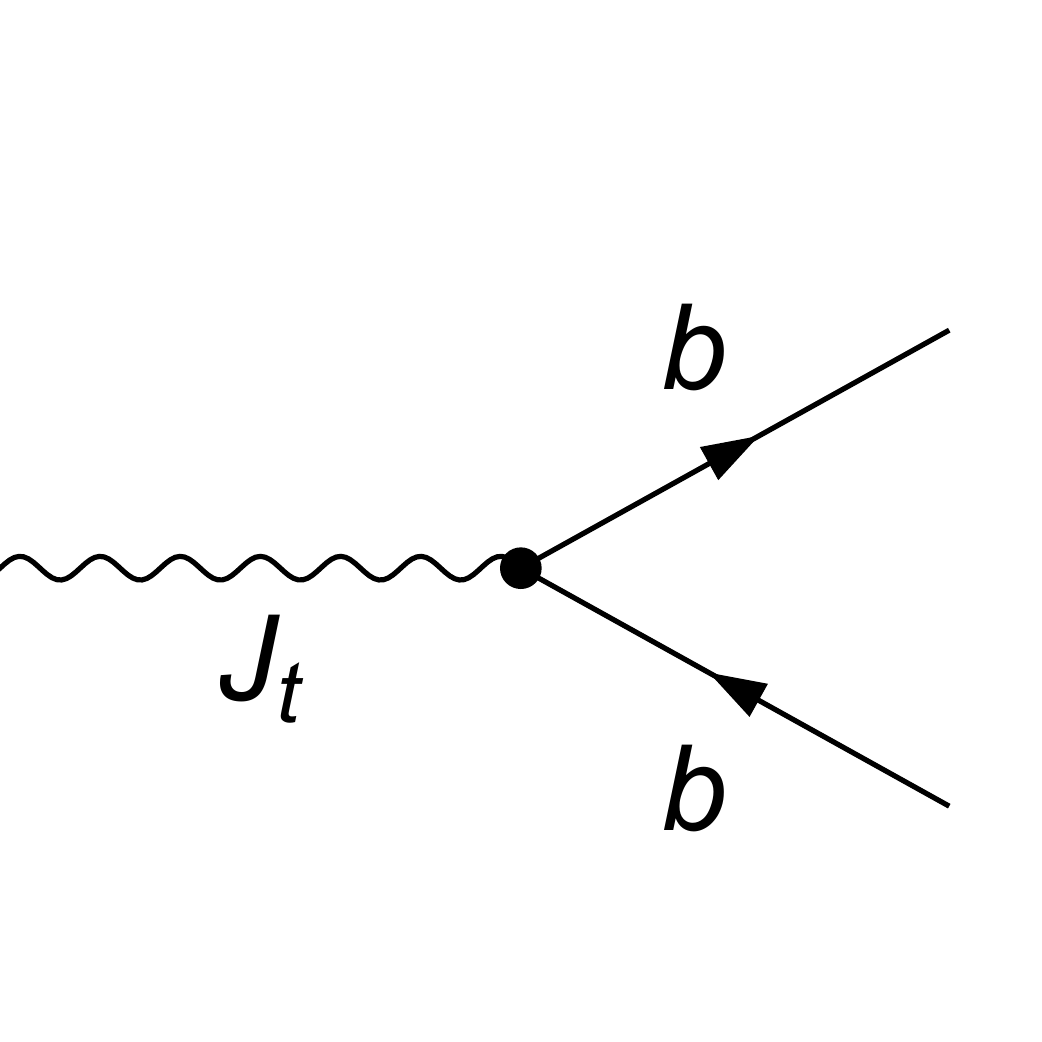}
    \end{subfigure}
    \begin{subfigure}[b]{0.3\textwidth}
        \includegraphics[width=\textwidth]{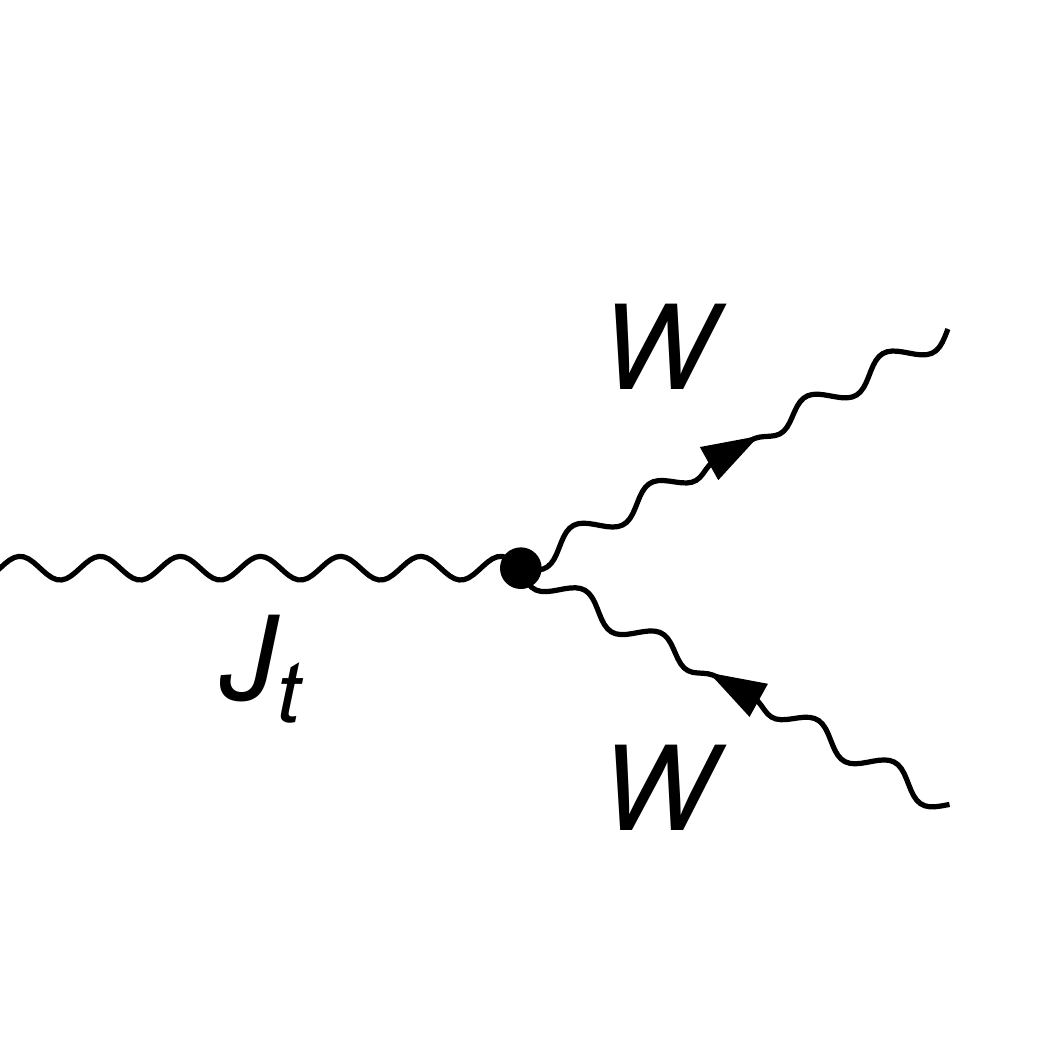}
    \end{subfigure}
    \caption{The two dominant Feynman diagrams for the decay processes of $J_t$}
    \label{fig:JtDecayFeynmanDiagram}
\end{figure}
\begin{eqnarray}
\mathcal{M}(J_t \to f\bar{f}) &=& -i\delta_{c1,c2} \bar{f}.\slashed{\varepsilon}(k_{J_t}).\left(g_{J_t ff}^V+g_{{J_t} ff}^V \gamma^5\right).f ,\nonumber\\
\mathcal{M}(J_t \to HZ) &=& ig_{{J_t} H Z}\left[k_H\cdot \varepsilon^*(k_Z)~k_H\cdot \varepsilon(k_{J_t})+\frac{m_H^2 \left(3 m_Z^2-4 m_t^2\right)+\left(m_Z^2-4 m_t^2\right){}^2}{8 m_t^2-2 m_Z^2}\varepsilon^*(k_Z)\cdot\varepsilon(k_{J_t})\right] ,\nonumber\\
\mathcal{M}(J_t \to HA) &=& -i g_{{J_t} HA}\left[k_H\cdot\varepsilon^*(k_A)~k_H\cdot \varepsilon(k_{J_t})+\frac{3}{2}\left(4m_t^2-m_H^2\right)\varepsilon^*(k_A)\cdot\varepsilon(k_{J_t})\right], \nonumber\\
\mathcal{M}(J_t \to Z_1 Z_2) &=& 2g_{{J_t} ZZ}~\epsilon_{\mu \nu \rho \sigma}\left(k_{Z_2}-k_{Z_1}\right)^\mu \varepsilon^{*\nu}(k_{Z_1}) \varepsilon^{*\rho}(k_{Z_2}) \varepsilon^\sigma(k_{J_t}), \nonumber\\
\mathcal{M}(J_t \to ZA) &=& 2g_{{J_t} ZA}~\epsilon _{\mu \nu \rho \sigma}{k_A}^\mu \varepsilon^{*\nu}(k_Z) \varepsilon^{*\rho}(k_A) \varepsilon^\sigma(k_{J_t}), \nonumber\\
\mathcal{M}(J_t \to W^+W^-) &=& i g_{{J_t} WW1}\left[k_{W^-}\cdot \varepsilon^{*}(k_{W^+})~\varepsilon^*(k_{W^-})\cdot \varepsilon(k_{J_t}) - k_{W^+}\cdot \varepsilon^{*}(k_{W^-})~\varepsilon^*(k_{W^+})\cdot \varepsilon(k_{J_t})\right] \nonumber\\
&&\hspace{1cm}-i g_{{J_t} WW2}~k_{W^+}\cdot \varepsilon(k_{J_t})~\varepsilon^*(k_{W^+})\cdot \varepsilon^*(k_{W^-}) \nonumber\\
&&\hspace{1cm}+g_{{J_t} WW3}~\epsilon_{\mu\nu\rho\sigma}(k_{W^+}-k_{W^-})^\mu \varepsilon^{*\nu}(k_{W^+}) \varepsilon^{*\rho}(k_{W^-}) \varepsilon^{\sigma}(k_{J_t}).
\end{eqnarray}
Here, $c1$ and $c2$ are the color indexes for $f$ and $\bar{f}$ respectively. The coupling constants are given by:
\begin{eqnarray}
g^A_{{J_t}ll}&=&g^{V}_{{J_t}\nu\nu}=-g^{A}_{{J_t}\nu\nu}=\sqrt{\frac{\pi \lambda ^3}{3}}\frac{\alpha  m_t^2 \left(3-8s_{\rm W}^2\right)}{4 s_{\rm W}^2 c_{\rm W}^2 \left(4 m_t^2-m_Z^2\right)},\nonumber \\
g^V_{{J_t}ll}&=&\sqrt{\frac{\pi \lambda ^3}{3}}\frac{\alpha  \left(8 m_W^2 s_{\rm W}^2-3 m_t^2 \left(1+4 s_{\rm W}^2 \right)\right)}{4 c_{\rm W}^2 s_{\rm W}^2 \left(4 m_t^2-m_Z^2\right)},\nonumber \\
g^A_{{J_t}uu}&=&g^A_{{J_t}cc}=-g^A_{{J_t}dd}=-g^A_{{J_t}ss}=\sqrt{\frac{\pi \lambda ^3}{3}}\frac{\alpha  m_t^2 \left(8 s_{\rm W}^2-3\right)}{4 c_{\rm W}^2 s_{\rm W}^2 \left(4 m_t^2-m_Z^2\right)},\nonumber \\
g^V_{{J_t}uu}&=&g^V_{{J_t}cc}=\sqrt{\frac{\pi \lambda ^3}{3}}\frac{\alpha  \left(m_t^2 \left(16 s_{\rm W}^2 +9\right)-16 m_W^2 s_{\rm W}^2\right)}{12 c_{\rm W}^2 s_{\rm W}^2 \left(4m_t^2-m_Z^2\right)},\nonumber \\
g^V_{{J_t}dd}&=&g^V_{{J_t}ss}=\sqrt{\frac{\pi \lambda ^3}{3}}\frac{\alpha  \left(m_t^2 \left(4 s_{\rm W}^2 -9\right)+8 m_W^2 s_{\rm W}^2\right)}{12 c_{\rm W}^2 s_{\rm W}^2 \left(4m_t^2-m_Z^2\right)},\nonumber \\
g^A_{{J_t}bb}&=&\sqrt{\frac{\pi \lambda ^3}{3}}\frac{\alpha  m_t^2 \left(\dfrac{3-8 s_{\rm W}^2}{c_{\rm W}^2 \left(4 m_t^2-m_Z^2\right)}+\dfrac{1}{m_t^2+m_W^2}+\dfrac{1}{m_W^2}\right)}{4 s_{\rm W}^2},\nonumber \\
g^V_{{J_t}bb}&=&-\sqrt{\frac{\pi \lambda ^3}{3}}\frac{\alpha}{12} \left(\frac{m_t^2 \left(\dfrac{4(8 c_{\rm W}^2+1)s_{\rm W}^2-9}{c_{\rm W}^2 \left(4 m_t^2-m_Z^2\right)}-\dfrac{3}{m_t^2+m_W^2}-\dfrac{3}{m_W^2}\right)}{s_{\rm W}^2}-8\right),\nonumber \\
g_{{J_t} HZ}&=& -\sqrt{\frac{\pi \lambda ^3}{3}}\frac{\alpha m_t^2(8s_{\rm W}^2-3)}{s_{\rm W}^2 c_{\rm W} m_W (4m_t^2-m_H^2-m_Z^2)},\nonumber \\
g_{{J_t} HA}&=& \sqrt{\frac{\pi \lambda ^3}{3}}\frac{8\alpha m_t^2}{s_{\rm W} m_W (4m_t^2-m_H^2)},\nonumber \\
g_{{J_t} ZZ}&=& -\sqrt{\frac{\pi \lambda ^3}{3}}\frac{\alpha m_t^2(8s_{\rm W}^2-3)}{4 s_{\rm W}^2 c_{\rm W}^2(2m_t^2-m_Z^2)},\nonumber \\
g_{{J_t} ZA}&=& \sqrt{\frac{\pi \lambda ^3}{3}}\frac{4\alpha m_t^2}{c_{\rm W}s_{\rm W}(4m_t^2-m_Z^2)},\nonumber \\
g_{{J_t} WW1}&=&\sqrt{\frac{\pi \lambda ^3}{3}}\frac{\alpha  \left(m_t^2 \left(m_Z^2 \left(8 s_{\rm W}^2-3\right)+12 m_W^2\right)-8 m_W^2 m_Z^2 s_{\rm W}^2\right)}{2 s_{\rm W}^2 \left(m_t^2-m_W^2\right)\left(4 m_t^2-m_Z^2\right)},\nonumber \\
g_{{J_t} WW2}&=&\sqrt{\frac{\pi \lambda ^3}{3}}\frac{4 \alpha  m_Z^2 s_{\rm W}^2 \left(m_W^2-m_t^2\right)+3 \alpha  m_t^2 \left(m_Z^2-2 \left(m_t^2+m_W^2\right)\right)}{s_{\rm W}^2\left(m_t^2-m_W^2\right) \left(4 m_t^2-m_Z^2\right)},\nonumber \\
g_{{J_t} WW3}&=&-\sqrt{\frac{\pi \lambda ^3}{3}}\frac{3\alpha m_t^2}{2s_{\rm W}^2(m_t^2-m_W^2)}.
\end{eqnarray}

The Lagrangian can be obtained as follows:
\begin{eqnarray}
\mathcal{L}_{J_t} &=& -\frac{1}{4} {J_t}^{\mu\nu} {J_t}_{\mu\nu} + \frac{1}{2} m_{J_t}^2 {J_t}^2, \nonumber\\
\mathcal{L}_{{J_t} ff} &=&\sum_{f=\nu_e,\nu_\mu,\nu_\tau,e,\mu,\tau,u,c,d,s,b} \bar{f}.\gamma_\mu.\left( g^V_{{J_t}ff}  + g^A_{{J_t}ff} \gamma^5 \right). f {J_t}^\mu, \nonumber\\
\mathcal{L}_{{J_t} HZ} &=& g_{{J_t} HZ} \left[\frac{m_H^2 m_Z^2}{4m_t^2 -m_Z^2}Z_\mu H+Z_{\mu \nu}(\partial^\nu H)\right]{J_t}^\mu ,\nonumber\\
\mathcal{L}_{{J_t} HA} &=&g_{{J_t} H A}~A_{\mu \nu}(\partial^\nu H){J_t}^\mu, \nonumber\\
\mathcal{L}_{{J_t} ZZ} &=& g_{{J_t} ZZ}~\varepsilon_{\mu\nu\rho\sigma} Z^{\mu \nu} Z^\rho {J_t}^\sigma ,\nonumber\\
\mathcal{L}_{{J_t} ZA} &=& g_{{J_t} ZA}~\varepsilon_{\mu\nu\rho\sigma} A^{\mu \nu} Z^\rho {J_t}^\sigma ,\nonumber\\
\mathcal{L}_{{J_t} WW} &=& \left[ i g_{{J_t} WW1} \left((\partial_\sigma W^-_\mu)W^{+\sigma}-(\partial_\sigma W^+_\mu)W^{-\sigma} \right)\right.\nonumber\\
&&\hspace{1cm}+\frac{i}{2}g_{{J_t} WW2}\left((\partial_\mu W^-_\sigma)W^{+\sigma}-(\partial_\mu W^+_\sigma)W^{-\sigma} \right) \nonumber\\
&&\hspace{1cm}\left.+g_{{J_t} WW3}~\varepsilon_{\mu\nu\rho\sigma}\left((\partial^\nu W^{-\rho})W^{+\sigma}-(\partial^\nu W^{+\sigma})W^{-\rho}\right) \right]{J_t}^\mu. 
\end{eqnarray}
The first term is the kinetic term and the mass term, ${J_t}^{\mu \nu}=\partial^\mu {J_t}^{\nu}-\partial^\nu {J_t}^{\mu}$ is the field strength tensor of ${J_t}$, and the remaining terms describe the interactions.

\section{\texorpdfstring{Toponium model of \texttt{FeynRules}}{Toponium model of FeynRules}}

\FeynRules~is a \Mathematica~package designed to compute Feynman rules based on specified particle properties and the Lagrangian. It provides a systematic framework for deriving Feynman rules directly from the Lagrangian, making it a powerful tool for studying new physics models. \FeynRules~supports the export of models in various formats, including \FeynArts~and UFO. The UFO format, in particular, is widely used in high-energy physics programs such as \MadGraph~and \WHIZARD~for event generation, simulations, and phenomenological studies \cite{Christensen:2008py,Alloul:2013bka,Alwall:2014hca}. Model files in \FeynRules~are identified by the \texttt{.fr} extension, which contains the necessary information about particles, parameters, and interactions. To incorporate new particle information and interaction Lagrangians, a new \texttt{.fr} file can be created and loaded alongside the SM file (\texttt{SM.fr}) in \FeynRules~to compute the corresponding Feynman rules. In the new \texttt{.fr} file, the following sections must be defined: \texttt{M\$ClassesDescription}, which specifies the particle content and their properties; \texttt{M\$Parameters}, which defines the model parameters; and the Lagrangian, which describes the interactions between particles. This modular approach allows for efficient implementation and validation of new physics models.

For $\eta_t$ and $J_t$, the information about the new particles must be provided in the \texttt{M\$ClassesDescription} section of the \texttt{.fr} file, which includes details such as particle names, spins, masses, and quantum numbers.
\begin{Verbatim}[frame=single, rulecolor=\color{black}]
M$ClassesDescription = {
S[343] == {
    ClassName        -> Et,
    SelfConjugate    -> True,
    Mass             -> {MEt, 341.021},
    Width            -> {WEt, 2.615},
    ParticleName     -> "Et",
    PDG              -> 5000002, 
    TeX              -> Subscript[E,t],
    PropagatorType   -> D,
    PropagatorArrow  -> None,
    PropagatorLabel  -> "Et",
    FullName         -> "Eta_t"
  },
V[343] == {
    ClassName        -> Jt,
    SelfConjugate    -> True,
    Mass             -> {MJt, 341.021},
    Width            -> {WJt, 2.604},
    ParticleName     -> "Jt",
    PDG              -> 5000001, 
    TeX              -> Subscript[J,t],
    PropagatorLabel  -> "Jt",
    PropagatorType   -> Sine,
    PropagatorArrow  -> None,
    FullName         -> "J_t"
  }
};
\end{Verbatim}
Additionally, the coupling constants and interaction terms must be specified in the \texttt{M\$Parameters} section, while the complete Lagrangian describing the interactions between particles should be included at the end of the file. This structured approach ensures that all necessary components of the model are clearly defined and can be processed by \FeynRules~to generate the corresponding Feynman rules.

For more details about the model, including specific implementation guidelines and examples, please refer to \url{https://github.com/fujinghang/Toponium}.

\section{Usage}

\subsection{Loading model}

Next, we generate the \FeynArts~model and UFO files using \FeynRules~at LO, and these files are essential for further phenomenological studies and event generation in tools like \MadGraph~and \WHIZARD. Before proceeding, we first set the directory of \FeynRules~to ensure that the package can correctly locate and load the necessary model files, such as \texttt{SM\_Updated.fr} and \texttt{Toponium.fr}. This step is crucial for the successful generation of Feynman rules and the subsequent export of model files.
\begin{mmaCell}{Input}
\$FeynRulesPath=NotebookDirectory[]<>"/feynrules-current";
<< \mmaDef{FeynRules}\`

\end{mmaCell}

For $\eta_t$ and $J_t$, we load the \texttt{SM\_Updated.fr} and \texttt{Toponium.fr} model files, along with the restriction files, and check the hermiticity of the Lagrangian. Here, \texttt{SM\_Updated.fr} contains the updated parameters~\cite{ParticleDataGroup:2024cfk,ATLAS:2024dxp,Proceedings:2019vxr,Herren:2017osy} for the SM:
{
\begin{eqnarray}
\label{eq:parameters}
\begin{array}{llll}  
&m_t=172.57~{\rm GeV},&~m_b=2.568~{\rm GeV},&~m_c=0.575~{\rm GeV},\\
&m_H=125.20~{\rm GeV},&~\alpha_s=0.09844,&~\alpha=1/126.04,\\
&m_Z=91.1880~{\rm GeV},&~m_W=80.3692~{\rm GeV},&~\cos \theta_{\rm W}=m_W/m_Z. 
\end{array} 
\end{eqnarray}
}
The coefficient $\lambda$ is fit from the center value of summed $\eta_t(nS)\to t \bar{t}$ cross section $8.8^{+1.2}_{-1.4}$~pb at 13 TeV at LHC measured by CMS~\cite{CMS:2025kzt}. With the parton distribution $\tt{NNPDF40\_lo\_ as\_01180 }$~\cite{Cruz-Martinez:2024cbz}  and 
\begin{eqnarray}
{\lambda=0.309}.
\end{eqnarray}
The cross sections for $pp \to \eta_t(1S)$ at various center-of-mass energies are presented in Table~\ref{tab:pptoEt}. Additionally, the production cross sections for $\eta_t(nS)$ states at $\sqrt{s} = 13~\mathrm{TeV}$ at the LHC are given by:
{
\begin{eqnarray}
\sigma(pp\to \sum_n \eta_t(nS))&=&7.36~\zeta(3)~{\rm pb}=8.85~{\rm pb},\nonumber \\
\sigma(pp\to \sum_n \eta_t(nS)\to {\rm non-}t\bar{t}~)&=&0.0647~\zeta(6)~{\rm pb}=0.0658~{\rm pb},\nonumber \\
\sigma(pp\to \sum_n \eta_t(nS)\to t\bar{t}~)&=&(8.85-0.0658)~{\rm pb}=8.78~{\rm pb}.
\end{eqnarray}
Here $\zeta(l)=\sum_n 1/n^l$ is Riemann Zeta function, and $\zeta(3)=1.20$,~$\zeta(6)=1.02$. The uncertainty of $\lambda$ is estimated as $\lambda=0.309\pm 0.010$, which is about $\alpha_s/\pi$ correction to $0.309$.}

\begin{table}[htbp]
    \centering
    \caption{LO cross sections (in pb) for the processes $pp \to \eta_t(1S)$ at various center-of-mass energies $\sqrt{s}$.}
    \label{tab:pptoEt}
    {
    \begin{tabular}{lcccccc}
    \toprule
       $\sqrt{s}$ (TeV)  & ~~~ 13 ~~~ &  13.6  &  14  &  27  &  50  &  100 \\
    \midrule
        $\sigma(pp \to \eta_t \to HZ)$ (pb)      
            & 0.0158   & 0.0176  & 0.0187  & 0.0741  & 0.230  & 0.708   \\
        $\sigma(pp \to \eta_t \to W^+ W^-)$ (pb) 
            &  0.00365 & 0.00405 & 0.00432 & 0.0170  & 0.0526 & 0.161   \\
        $\sigma(pp \to \eta_t \to gg)$ (pb)      
            &  0.0447  & 0.0495  & 0.0529  & 0.208   & 0.641  & 1.95   \\
        $\sigma(pp \to \eta_t \to non-t\bar{t})$ (pb) 
            &  0.0647  & 0.0716  & 0.0765  & 0.301   & 0.929  & 2.83 \\ 
        $\sigma(pp \to \eta_t \to t\bar{t})$ (pb) 
            &  7.30    & 8.08    & 8.63    & 34.0    & 105.   &  320.  \\
        $\sigma(pp \to \eta_t)$ (pb)            
            &  7.36    &  8.15   &  8.70   & 34.3    & 106.   &  323.  \\
    \bottomrule
    \end{tabular}
    }
\end{table}

While \texttt{Toponium.fr} includes the toponium model with the necessary particle definitions and interactions. The restriction files, such as \texttt{Cabibbo.rst}, \texttt{Massless.rst}, and \texttt{DiagonalCKM.rst}, enforce specific constraints: \texttt{Cabibbo.rst} restricts the Cabibbo angle, \texttt{Massless.rst} ensures certain particles remain massless, and \texttt{DiagonalCKM.rst} imposes a diagonal CKM matrix. To verify the hermiticity of the Lagrangian, \texttt{CheckHermiticity} calculates the Feynman rules derived from $\texttt{L-HC[L]}$, where $\texttt{HC[L]}$ denotes the Hermitian conjugate of the Lagrangian ($\texttt{L}$). If the Lagrangian is Hermitian, the number of vertices generated by $\texttt{L-HC[L]}$ should be zero, confirming the consistency of the model.
\begin{mmaCell}{Input}
SetDirectory[\mmaDef{\$FeynRulesPath}<>"/Models/Toponium"];
\mmaDef{LoadModel}["SM_Updated.fr","Toponium.fr"]
\mmaDef{LoadRestriction}["Cabibbo.rst","DiagonalCKM.rst"]
\mmaDef{CheckHermiticity}[LSM+L1Et+L1Jt]

\end{mmaCell}
Here, $\texttt{LSM}$ represents the Lagrangian of the SM, which includes all the known interactions between SM particles. $\texttt{L1Et}$ and $\texttt{L1Jt}$ denote the additional Lagrangian terms for the new particles $\eta_t$ and $J_t$, respectively. These terms describe the interactions of $\eta_t$ and $J_t$ with the SM particles. By combining $\texttt{LSM}$, $\texttt{L1Et}$, and $\texttt{L1Jt}$, we obtain the complete Lagrangian of the extended model, which can be used to derive the corresponding Feynman rules and study the phenomenological implications of the new particles.

\subsection{Decay width}

The partial decay widths of $\eta_t$ and $J_t$ are computed using the \texttt{ComputeWidth} function within the \FeynRules~framework, which automates the derivation of decay amplitudes through symbolic manipulation of the model Lagrangian. The corresponding implementation is provided in the following code snippet:
\begin{mmaCell}{Input}
\mmaDef{ComputeWidths[FeynmanRules[L1Et+L1Jt]]};
\%[[Position[\%[[All,1,1]],\mmaDef{Et}]//Flatten]];
\{\%[[All,1]],
  \%[[All,2]]//\mmaDef{NumericalValue}\}
\%\%\%[[Position[\%\%\%[[All,1,1]],\mmaDef{Jt}]//Flatten]];
\{\%[[All,1]],
  \%[[All,2]]//\mmaDef{NumericalValue}\}
 
\end{mmaCell}

The two-body annihilation decay widths and branching ratios of $\eta_t$ and $J_t$ are calculated and summarized in Table~\ref{tab:DecayWidthFR}. The total decay widths are set to {$\Gamma_{\eta_t} = 2.615~\text{GeV}$} and {$\Gamma_{J_t} = 2.604~\text{GeV}$} for $\eta_t$ and $J_t$, respectively.
\begin{eqnarray}
    \Gamma_{\eta_t/J_t}=2\Gamma_t \left(1-\frac{\lambda^2}{8}\right)+\Gamma^{\rm Anni.}_{\eta_t/J_t}.
\end{eqnarray}
Where $\Gamma_t=(1.3148^{+0.003}_{-0.005}+0.027(m_t-172.69))~{\rm GeV}$~\cite{chen2023topquarkdecaynexttonexttonexttoleadingorder,Chen:2023dsi}, and the relativistic correction factor $(1 - \lambda^2/8)$ originates from the distinction between bound-state top quarks in the $t\bar{t}$ system and free top quarks~\cite{Uberall:1960zz}.
We observe that the dominant two-body annihilation decay channels are $\eta_t \to gg/ZH$ for $\eta_t$ and $J_t \to W^+W^-/b\bar{b}$ for $J_t$, as indicated by their large branching ratios.
\begin{table}[htbp]
    \centering
    \caption{Partial widths $\Gamma_{\eta_t\to XY}$ and branching ratios ${\cal B}$ of $\eta_t$ and $J_t$ computed using \FeynRules.}
    \label{tab:DecayWidthFR}
    {
    \begin{tabular}{crrrr}
    \toprule
        Decay Channels & $\Gamma_{\eta_t\to XY}$~(MeV) & ${\cal B}_{\eta_t}~(10^{-4})$ & $\Gamma_{J_t\to XY}$~(MeV) & ${\cal B}_{J_t}~(10^{-4})$\\
    \midrule
        $W^+W^-$                        & 1.302  & 4.98  & 5.466 & 20.99 \\
        $b\bar{b}$                      & 0.0    & 0.0   & 4.914 & 18.87 \\
        $ZA$                            & 0.023  & 0.09  & 0.703 & 2.70 \\
        $HA$                            & 0.0    & 0.0   & 0.609 & 2.34 \\ 
        $HZ$                            & 5.624  & 21.51 & 0.081 & 0.31 \\
        $ZZ$                            & 0.067  & 0.25  & 0.055 & 0.21 \\
        $gg$                            & 15.864 & 60.67 & 0.0   & 0.0 \\
        $AA$                            & 0.092  & 0.35  & 0.0   & 0.0 \\
        $u\bar{u}/c\bar{c}$             & 0.0    & 0.0   & 0.134 & 0.52 \\
        $d\bar{d}/s\bar{s}$             & 0.0    & 0.0   & 0.064 & 0.25 \\
        $e^+e^-/\mu^+\mu^-/\tau^+\tau^-$
                                        & 0.0    & 0.0   & 0.078 & 0.30 \\
        $\nu_e\bar{\nu}_e/\nu_\mu\bar{\nu}_\mu/\nu_\tau\bar{\nu}_\tau$   
                                        & 0.0    & 0.0   & 0.008 & 0.03 \\
    \midrule\midrule
       Two-body Annihilation Decay      & 23     & 88    & 12    & 48 \\
       $t\bar{t}\to W^+b W^-\bar{b}$    & 2592   & 9912  & 2592  & 9952 \\
       Total                            & 2615   & 10000 & 2604  & 10000 \\
    \bottomrule
    \end{tabular}
    }
\end{table}

\subsection{\texorpdfstring{Output \texttt{FeynArts}~model and UFO files}{Output FeynArts model and UFO files}}

The Lagrangian is loaded, and the corresponding \FeynArts~model files are generated as follows:
\begin{mmaCell}{Input}
\mmaDef{FeynmanGauge}=True;
\mmaDef{WriteFeynArtsOutput}[
  \mmaDef{L1JtZA},\mmaDef{LSM}+\mmaDef{L1Et}+\mmaDef{L1Jt}-\mmaDef{L1JtZA},
  \mmaDef{Output}->NotebookDirectory[]<>"FeynArts-3.12/Models/Toponium_LO_FA",
  \mmaDef{FlavorExpand}->True]

\end{mmaCell}
Here, the \texttt{L1JtZA} term must be separated from \texttt{L1Jt} due to an unknown bug in the implementation.

Subsequently, UFO files are generated as follows:
\begin{mmaCell}{Input}
\mmaDef{WriteUFO}[\mmaDef{LSM}+\mmaDef{L1Et}+\mmaDef{L1Jt},\mmaDef{Output}->"Toponium_LO_UFO"]

\end{mmaCell}

\section{Example}

\subsection{\texorpdfstring{Example Using \texttt{FeynCalc}}{Example Using FeynCalc}}

\FeynCalc~is a \Mathematica~package designed for the symbolic evaluation of Feynman diagrams and algebraic calculations in quantum field theory and elementary particle physics~\cite{Shtabovenko:2023idz,Shtabovenko:2020gxv,Shtabovenko:2016sxi,Mertig:1990an}. Since \FeynArts~models are not directly compatible with \FeynCalc, it is necessary to replace certain variable names using the following command:
\begin{verbatim}
    FAPatch[PatchModelsOnly->True]
\end{verbatim}
\texttt{FAPatch} is a function that patches \FeynArts~to be compatible with \FeynCalc.

Next, we proceed to calculate specific processes using \FeynCalc. Before performing any calculations, \FeynCalc~must be installed on the system. Once installed, it can be loaded using the following command:
\mmaSet{index=1}
\begin{mmaCell}{Input}
\$LoadFeynArts = True;
<< \mmaDef{FeynCalc}\`

\end{mmaCell}

\begin{figure}[htbp]
    \centering
        \includegraphics[width=0.3\textwidth]{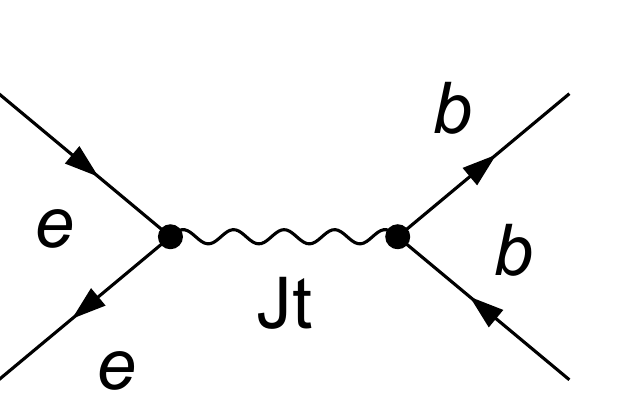}
    \caption{The Feynman diagram for the process of $e^-e^+ \to J_t \to b \bar{b}$ in an electron-positron collider.}
    \label{fig:eeJtbbFeynmanDiagram}
\end{figure}
For example, to calculate the Born cross section for the process $e^+e^- \to J_t \to b\bar{b}$, the following code can be used:
\begin{mmaCell}{Input}
\mmaDef{ClearScalarProducts};
\mmaDef{SetMandelstam}[s,t,u,pe1,pe2,-k1,-k2,0,0,0,0];
\mmaDef{InsertFields}[\mmaDef{CreateTopologies}[0,2->2],
  \{\mmaDef{F}[4],-\mmaDef{F}[4]\}->\{\mmaDef{F}[12],-\mmaDef{F}[12]\},
  \mmaDef{InsertionLevel}->\{\mmaDef{Particles}\},\mmaDef{ExcludeParticles}->\{\mmaDef{V}[4]\},\mmaDef{LastSelections}->\{\mmaDef{V}[5]\},
  \mmaDef{Model}->"Toponium_LO_FA",\mmaDef{GenericModel}->"Toponium_LO_FA"];
\mmaDef{FCFAConvert}[\mmaDef{CreateFeynAmp}[\%],
  \mmaDef{IncomingMomenta}->\{pe1,pe2\},\mmaDef{OutgoingMomenta}->\{k1, k2\},List->False]//.
  \mmaDef{FCGV}[xx_]:>ToExpression[xx]//.
  \mmaDef{FeynAmpDenominator}[\mmaDef{PropagatorDenominator}[\mmaDef{Momentum}[x__],MJt]]->
  \mmaDef{FeynAmpDenominator}[\mmaDef{PropagatorDenominator}[\mmaDef{Momentum}[x],Sqrt[MJt^2-I*WJt*MJt]]]//
  \mmaDef{FeynAmpDenominatorExplicit}//\mmaDef{Contract}//Simplify;
1/(64*Pi^2*s)*2 Pi*1/4*(\%*\mmaDef{ComplexConjugate}[\mmaDef{FCRenameDummyIndices}[\%]]//
  \mmaDef{FermionSpinSum}//\mmaDef{Calc})//.\mmaDef{CA}->3//.\{u->-s-t, t->1/2(costh-1)*s\}//.
  \{Me->0,MB->0\}//.\mmaDef{M\$FACouplings}//Integrate[#,\{costh,-1,1\}]&

\end{mmaCell}

The Feynman diagram is shown in Fig.~\ref{fig:eeJtbbFeynmanDiagram}. The LO cross section of $e^+e^- \to J_t \to b\bar{b}$ can be obtained as follows:
\begin{eqnarray}
    \sigma^{\rm Born}(e^+e^- \to J_t \to b\bar{b})=\frac{s\left({g_{J_t bb}^{A}}^2+{g_{J_t bb}^{V}}^2\right)\left({g_{J_t ll}^{A}}^2+{g_{J_t ll}^{V}}^2\right)}{4\pi \left[\left(s-m_{J_t}^2\right)^2+m_{J_t}^2 \Gamma_{J_t}^2\right]}\xlongequal{\sqrt{s}=m_{J_t}}
    {7.12}~{\rm fb}.
\end{eqnarray}
Detailed results, related background studies, and uncertainties analyses for the process $e^+e^- \to J_t \to b\bar{b}$ are provided in Ref.~\cite{Fu:2024bki}.

\begin{figure}[htbp]
    \centering
        \includegraphics[width=0.3\textwidth]{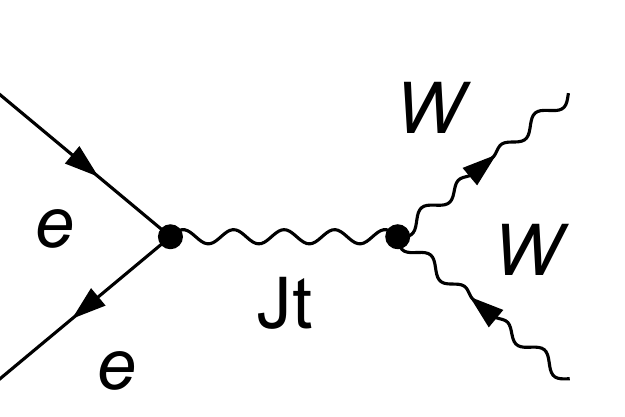}
    \caption{The Feynman diagrams for the process of $e^-e^+ \to J_t \to W^+ W^-$ in an electron-positron collider.}
    \label{fig:eeJtWWFeynmanDiagram}
\end{figure}

Similarly, the Feynman diagram of the process $e^+e^- \to J_t \to W^+W^-$  is shown in Fig.~\ref{fig:eeJtWWFeynmanDiagram}, and the LO cross section is:
\begin{eqnarray}
    \sigma(e^+e^- \to J_t \to W^+W^-)=
    {7.92}~{\rm fb}.
\end{eqnarray}

\subsection{\texorpdfstring{Example Using \texttt{MadGraph5\_aMC@NLO}}{Example Using MadGraph5\_aMC@NLO}}

For example, we can calculate the decay width and cross section using \texttt{MadGraph5\_aMC@NLO}. This tool enables precise computations of physical observables at both LO and NLO in perturbative expansions. To proceed, place the UFO model files into the directory \texttt{MG5\_aMC\_v3\_x\_x/models}.

Next, navigate to the root directory of \texttt{MadGraph5\_aMC@NLO} and create a text file named \texttt{test\_pp\_et+jet.txt} containing the following code for the process $pp \to \eta_t + \text{jet}$:
\begin{Verbatim}[frame=single, rulecolor=\color{black}]
    import model Toponium_LO_UFO/
    generate p p > et j
    output PROC_pp_et+jet
    launch
    0
    set ebeam1 6500
    set ebeam2 6500
    set pdlabel lhapdf
    set lhaid 331900
    set ptj 150
    0
    launch
    0
    set ebeam1 6800
    set ebeam2 6800
    0
    ...
    ...
\end{Verbatim}
Here, we employ the parton distribution function \texttt{NNPDF40\_lo\_as\_01180}~\cite{Cruz-Martinez:2024cbz}. We use this script to generate the corresponding process:
\begin{Verbatim}[frame=single, rulecolor=\color{black}]
    $ ./bin/mg5_aMC test_pp_et+jet.txt
\end{Verbatim}

Similarly, we can also calculate the cross section for the process $pp \to J_t + \text{jet}$. The LO cross section results for both processes at various center-of-mass energies are presented in Table~\ref{tab:MG5Examplepp}. These results provide valuable input for future higher-order calculations and experimental searches at hadron colliders.
\begin{table}[htbp]
    \centering
    \caption{LO cross sections (in pb) for the processes $pp \to \eta_t + \mathrm{jet}$ and $pp \to J_t + \mathrm{jet}$ at various center-of-mass energies $\sqrt{s}$, with a minimum jet transverse momentum requirement $p_T^\mathrm{jet} > 150~\mathrm{GeV}$.}
    \label{tab:MG5Examplepp}
    {
    \begin{tabular}{ccc}
    \toprule
       $\sqrt{s}$ (TeV)  & ~~~$\sigma(pp \to \eta_t +{\rm jet})$ (pb)~~~ & ~~~$\sigma(pp \to J_t +{\rm jet})$ (pb)~~~  \\
    \midrule
        13   & 0.972 & 0.00231 \\
        13.6 & 1.11  & 0.00254 \\
        14   & 1.20  & 0.00268 \\
        27   & 6.19  & 0.00893 \\
        50   & 23.7  & 0.0241 \\
        100  & 90.0  & 0.0655 \\
    \bottomrule
    \end{tabular}
    }
\end{table}

As a next-generation collider concept, muon colliders provide unique capabilities for high-energy precision physics through their characteristic low-background environment and exceptional energy reach, enabled by the muon's significant mass ($m_\mu \approx 106$ MeV) and negligible synchrotron radiation losses~\cite{Accettura:2023ked,Han:2024gan,InternationalMuonCollider:2024jyv,Han:2025wdy}. This work will also serve as a valuable reference for future experimental  and theoretical studies at electron-positron colliders.

\section{Summary and outlook}
In this work, we have developed a comprehensive framework for studying the production and decay mechanisms of toponium ($\eta_{t}$ and $J_{t}$) within the SM model using FeynRules. By incorporating NRQCD and a Coulomb potential approximation, we derived the effective Lagrangians for the spin-singlet $\eta_{t}$ and spin-triplet $J_{t}$ bound states. The couplings of these toponium states to SM particles were systematically calculated, enabling the generation of FeynArts, UFO, and WHIZARD model files. These models facilitate efficient simulations of toponium processes in collider experiments.

Our key results include the decay widths and branching ratios of $\eta_{t}$ and $J_{t}$, as well as LO cross sections for their production in $pp$ and $e^{+}e^{-}$ collisions. For example, the decay width of $\eta_{t}$ to $gg$ channel is found to be consistent with the theoretical expectations within a {5\%} error margin, which is a significant validation of our model. In addition, the LO cross section for the two-body annihilation decay part of  $pp \to \eta_{t}(nS) $ is about {0.06 pb}  at 13 TeV, providing a crucial benchmark for experimental searches.

Furthermore, this framework not only provides a powerful tool for studying toponium properties but also offers new insights into the exploration of new physics scenarios beyond the SM.
Notably, the spin correlations in \(J_t(\eta_t)\to W^+W^-\)  decays could serve as a probe of quantum entanglement, offering a novel pathway to test quantum foundations in collider experiments. Additionally, deviations in \(\eta_t\to ZH\) branching ratios may signal Higgs portal interactions beyond the SM.
The precise calculations of toponium processes can help us better understand the fundamental interactions and potentially discover new particles or phenomena. For instance, the deviations in the production and decay processes of toponium from the SM predictions might indicate the existence of new physics, such as extra dimensions or new particles that interact with the top-quark sector.

Looking ahead, with the development of more advanced collider technologies like the future muon colliders and the upgrade of existing facilities, we expect to obtain more precise experimental data. This will further test and refine our model. We plan to extend our study to include higher-order corrections, such as electroweak corrections, which were neglected in this work but could be significant in more precise calculations. Additionally, we aim to explore the implications of our model in different energy scales and collider environments to gain a more comprehensive understanding of toponium physics.

\section*{Acknowledgements}

We thank Yu-Ji~Li and Profs. Cheng-Ping~Shen, and Li Yuan for valuable and helpful discussions. This work is supported by National Natural Science Foundation of China (NSFC) under contract No.~12075018 and No.~11705078.











\end{document}